\documentclass[letter]{aa}

\usepackage{graphicx}
\usepackage{txfonts}
\usepackage{caption}
\usepackage{subcaption}
\usepackage{siunitx}
\usepackage{bm}
\usepackage{hyperref}
\hypersetup{
    colorlinks,
    linkcolor={black},
    citecolor={blue!50!black},
    urlcolor={blue!80!black}
}
\usepackage{physics}

\makeatletter

\begin{document}

   \title{Asteroseismic detection of an internal magnetic field in the B0.5V pulsator HD\,192575}

  \author{Jelle Vandersnickt \inst{1}
            \and 
            Vincent Vanlaer \inst{1}
            \and 
            Mathijs Vanrespaille \inst{1}
            \and 
            Conny Aerts \inst{1, 2, 3}
          }

   \institute{Institute of Astronomy, KU Leuven, Celestijnenlaan 200D, 3001, Leuven, Belgium\\
              \email{jelle.vandersnickt@kuleuven.be}
              \and 
              Department of Astrophysics, IMAPP, Radboud University Nijmegen, PO Box 9010, 6500 GL Nijmegen, The Netherlands
               \and
               Max Planck Institute for Astronomy, Königstuhl 17, 69117 Heidelberg, Germany
         }

   \date{Received, Accepted}

  \abstract
{Internal magnetic fields are an elusive component of stellar structure. However, they can play an important role in stellar structure and evolution models through efficient angular momentum transport and their impact on internal mixing. }
{We strive to explain the 9 components of one frequency multiplet, identified as a low-order quadrupole gravity mode detected in the light curve of the $\beta$\,Cep pulsator HD\,192575 assembled by the Transiting Exoplanet Survey Satellite (TESS).}
{We update the frequencies of the quadrupole mode under investigation using a standard prewhitening method applied to the 1951.46\,d TESS light curve, showing that an internal magnetic field is required to simultaneously explain all 9 components. We implement theoretical pulsation computations applicable to the low-order modes of a $\beta$\,Cep pulsator
including the Coriolis force, as well as a magnetic field that is misaligned with respect to the rotation axis. We apply the theoretical description 
to perform asteroseismic modelling of the amplitudes and frequencies in the multiplet of the quadrupole g-mode of this evolved $\beta\,$Cep star.}
{Pulsation predictions based on the measured internal rotation frequency of the star cannot explain the observed 9-component frequency splittings of the quadrupole low-order gravity mode. By contrast, we show that the combined effect of the Coriolis force caused by the near-core rotation with a period of $\sim\! 5.3$\,d and the Lorentz force  due to an internal inclined magnetic field with a maximum strength of $\sim\! 24$\,kG does provide a proper explanation of the 9 multiplet frequencies and their relative amplitudes.}
{Given HD\,192575's stellar mass of about 12\,M$_\odot$, this work presents the detection and magneto-gravito-asteroseismic modelling of a stable internal magnetic field buried inside an evolved rotating supernova progenitor. }


   \maketitle
%

\section{Introduction}

Measured internal rotation rates of intermediate-mass stars are lower than expected from standard stellar evolution theories, leading to a debate about the theory of angular momentum transport \citep{aertsAngularMomentumTransport2019a,AertsTkachenko2024}. Moreover, near-uniform rotation was reported for main-sequence stars with masses between $1.3$\,M$_\odot$ and $3.3$\,M$_\odot$ \citep{aertsProbingInteriorPhysics2021a} 
and low levels of differential rotation between 1 and 2 occur for the $\beta\,$Cep pulsators
\citep[8\,M$_\odot\!\leq$M$\leq$25\,M$_\odot$,][]{Fritzewski2025}. These asteroseismic results
indicate an angular momentum transport problem, where the transport occurs more efficiently than predicted already early on in the lives of stars. 

Along with internal gravity waves, internal magnetic fields are expected to efficiently transport angular momentum. 
Such field could possibly also explain the near-uniform rotation rates of 
stars in core-hydrogen or core-helium burning stages \citep{aertsAngularMomentumTransport2019a}.
While surface magnetic fields have been widely detected for most types of stars \citep{landstreetMagneticFieldsSurfaces1992,donatiMagneticFieldsNondegenerate2009}, internal fields are notoriously difficult to detect and characterize.
The internal magnetic fields in the radiative envelope of intermediate- and high-mass stars are thought to be of fossil origin \citep{mestelStellarMagnetism1999, neinerOriginMagneticFields2015}, where the field is a remnant of a primordial magnetic field present during the star formation process.

Asteroseismology provides a unique and promising opportunity to study internal magnetic fields buried beneath the surface by probing stellar interiors through nonradial oscillations. 
This has been achieved from ground-based asteroseismology for the strongly magnetic rapidly oscillating Ap (roAp) stars \citep{Kurtz1982} and led to the oblique pulsator model \citep{Shibahashi1993,Bigot2002}. This model was also successfully applied to explain
the multiplet behaviour of the dominant radial mode in the prototypical star $\beta\,$Cephei \citep{Shibahashi2000}. Since the space asteroseismology era, 
internal magnetic fields were detected in red giants \citep{liMagneticFields302022a,liInternalMagneticFields2023a,deheuvelsStrongMagneticFields2023a,hattAsteroseismicSignaturesCore2024} and interpreted by
the theoretical frameworks developed by \citet{gomesCoreMagneticField2020}, \citet{mathisProbingInternalMagnetism2021}, \citet{loiTopologyObliquityCore2021a}, \citet{liMagneticFields302022a}, \citet{bugnetMagneticSignaturesMixedmode2022b}, and \citet{mathisAsymmetriesFrequencySplittings2023}.
Aside from direct magnetically induced frequency detections in roAp stars and $\beta\,$Cephei, the presence of internal magnetic fields in 
intermediate- or high-mass main-sequence stars is
indirect and based on theory \citep{Aerts2021-GIW,Rui2024}, numerical simulations 
\citep{lecoanetAsteroseismicInferenceNearcore2022a}, or inferences
from depressed dipole modes in red giants
\citep{fullerAsteroseismologyCanReveal2015,2016Natur.529..364S}.

This Letter reports the direct detection as well as asteroseismic modelling of an internal magnetic field in a $\sim$12\,M$_\odot$ evolved main-sequence star from
one of its low-order pulsation modes, which occurs far outside the asymptotic regimes of low or high frequencies
\citep[see Fig.\,3 in][]{aertsAsteroseismicModellingFast2024}.

\vspace{-3pt}
\section{Observational input and equilibrium models}\label{sec:observations}
\vspace{-3pt}
HD\,192575 is an early-type B0.5V star discovered to be a $\beta\,$Cep pulsator by \citet{burssensCalibrationPointStellar2023b}. They analysed its 324\,d light curve
obtained in Cycle\,2 of the Transiting Exoplanet Survey Satellite \citep[TESS,][]{rickerTransitingExoplanetSurvey2015}
and deduced a rich oscillation spectrum with several multiplets of low-order low-degree pressure and gravity modes.
The authors additionally derived an effective temperature of $T_{\rm eff} = 23\,900 \pm 900\,{\rm K}$, a surface gravity of $\log g = 3.65 \pm 0.15$, and the rotation velocity projected onto the line-of-sight $v \sin i=27 ^{+6}_{-8}$\,km/s from spectra.

From stellar evolution models computed with
the \texttt{MESA} software \citep[Modules for Experiments in Stellar Astrophysics, Version 12778,][]{paxtonModulesExperimentsStellar2019} 
HD\,192575 was found to have a mass between 11 and 13\,M$_\odot$. The star is in an evolved stage approaching the end of the main sequence as $X\leq 0.25$ in its convective core. This was derived from extensive forward seismic modelling by \citet{Vanlaer2025}, who used the Cycles\,2, 4, and 5 TESS light curve. From rotation inversions, this study also led to a well-constrained rotation frequency of about 0.2\,d$^{-1}$ in the zone adjacent to the convective core and a ratio of the core-to-surface rotation between 0.4 and 1.8. This is in agreement with the result of 1.4 obtained from two-dimensional evolution models computed with the code {\tt ESTER} by \citet{mombargTwodimensionalPerspectiveRotational2024}.

Following TESS Cycles\,2, 4,  5, and 6, HD\,192575 has
a 2-minute cadence light curve covering 1951.46\,d (Rayleigh frequency resolution of 
$\sim\SI{5e-4}{\per\day}$) with a duty cycle of 43\%.
Figure~\ref{ChIn:fig-HD192575-lightcurveApp} shows its 
Lomb-Scargle periodogram.
We performed a new frequency analysis to extract periodic signal in this light curve as an improvement to the results in \citet{burssensCalibrationPointStellar2023b} and \citet{Vanlaer2025} with the addition of the data of Cycle\,6 (up to December 2024).
We applied classical prewhitening, extracting the dominant frequency at each step, optimising its value, amplitude, and phase from nonlinear regression, and computing a residual curve for the next iteration. This was done by an adapted version of the \texttt{STAR\,SHADOW} software \citep{ijspeertAutomatedEccentricityMeasurement2024}. All the extracted frequencies with a local signal-to-noise ratio (SNR) above 2 are provided electronically in Table\,A.1 in Appendix\,A, where the SNR is calculated as the ratio of the extracted amplitude to the mean amplitude in a window of 1\,$\text{d}^{-1}$ centered on the frequency of interest. 
Each frequency in the nine-component multiplet is given an identifier $f^j_g$ where the index $j$ increases from low to high frequencies.
The new data lead to the additional detection of $f^2_g$ compared to \citet{Vanlaer2025} and $f^2_g$ and $f^9_g$ compared to \citet{burssensCalibrationPointStellar2023b}.

For the purpose of this Letter we rely on the forward models that fit the 15 identified frequencies within the 4 multiplets selected and identified by 
\citet{Vanlaer2025}. We dedicatedly 
focus on the multiplet of the quadrupole gravity (g-)mode at 3.8\,d$^{-1}$ 
as the only multiplet in the entire oscillation spectrum revealing more frequencies than can be explained by rotational splitting of a single mode.
Table~\ref{ChFit-tab:fitting_results} lists the 9 extracted frequencies, amplitudes, and SNR of this multiplet, which is
shown in Fig.\,\ref{ChIn:fig-HD192575-lightcurve}.
Based on the 7 available frequencies from Cycle\,2, \citet{burssensCalibrationPointStellar2023b} interpreted the selected g-mode multiplet as resulting from two merged quadrupole modes undergoing an avoided crossing \citep{aizenmanAvoidedCrossingModes1977}. 
However, the more extended light curve used by \citet{Vanlaer2025} revealed that the avoided crossing assumption is invalid, unless the spectroscopic error box is enlarged to 3$\sigma$ in $T_{\rm eff}$ and $\log g$. Because of this, \citet{Vanlaer2025}  selected additional model sets fitting the 15 identified frequencies, among which the quadrupole low-order g modes with the five frequencies
$f^1_g$, $f^4_g$, $f^6_g$, $f^7_g$, and $f^9_g$. The authors  assumed these 
to form a complete rotationally split asymmetric quintuplet, leaving the other frequencies in Table~\ref{ChFit-tab:fitting_results}  
and Fig.\,\ref{ChIn:fig-HD192575-lightcurve}
unexplained.
The additional detected oscillation frequency in our analysis, $f^2_g$, decreased the maximum spacing between the two adjacent modes in the avoided crossing from \SI{0.18}{\per\day} to \SI{0.02}{\per\day}.
None of the stellar models in the model grid of \citet{burssensCalibrationPointStellar2023b} 
used by
\citet{Vanlaer2025} allow for such a constraint, irrespective of any spectroscopic cut, firmly excluding the interpretation as an avoided crossing.
The new data thus leave only an internal magnetic field as unexplored avenue to interpret all 9 observed frequencies simultaneously as spectroscopy revealed the star to be single.
Hence, we included the effect of a magnetic field in a theoretical pulsation model, which is capable of explaining all the nine frequencies in Table~\ref{ChFit-tab:fitting_results} and Fig.\,\ref{ChIn:fig-HD192575-lightcurve} from models within the $1\sigma$ spectroscopic box of the star fitting the 15 selected identified multiplet frequencies considered by 
\citet{Vanlaer2025}.

\vspace{-9pt}
\section{Theoretical pulsation approximation}\label{sec:model}
\vspace{-3pt}
We calculated the effects of rotation and an internal magnetic field on the oscillation modes of the star. 
To do so, we considered these effects in a first-order approximation ignoring the centrifugal force. This is justified given the modest rotation rate and the mode under study having a spin parameter of $\sim$\!10\%. 
Moreover, the quadrupole g-mode's rotation kernel is dominant near the convective core for most models.
We assumed solid body rotation as a proxy for the average rotation rate in the region where the g-mode probes.
Under these circumstances we can rely on 
the theoretical framework put forward by \citet{loiTopologyObliquityCore2021a} for red giants to calculate the relative amplitudes and frequencies of the oscillation modes. We did so for the 6 model sets in \citet{Vanlaer2025}, whose properties are repeated in Appendix\,A, Table\,\ref{ChFit-tab:stellar_models} for convenience.
 
The eigenfrequency $\omega_0$ for the corresponding eigenvector $\bm{\xi}$ of an oscillation mode in the absence of rotation and internal magnetic fields follows from the equation \citep{1990MNRAS.242...25G, aertsAsteroseismology2010}

\begin{equation}
    \omega^2_0 \bm{\xi} = \vb{\mathcal{F}_0} \left( \bm{\xi} \right) = \frac{1}{\rho_0} \grad{p'} - \vb{g'} - \frac{\rho'}{\rho_0}\vb{g_0},
\end{equation}
describing the mode as a consequence of linear perturbations in the density $\rho$, pressure $p$, and the gravitational acceleration $\bm{g}$.
Following \citet{loiTopologyObliquityCore2021a}, we included the effects of rotation and a misaligned internal magnetic field by perturbing the equation of motion and obtain 
\begin{equation}
    (\omega^2_0 + \delta \omega^2) \bm{\xi} = \vb{\mathcal{F}_0} \bm{\xi} + \vb{\delta \mathcal{F}_{\rm rot}} \bm{\xi} + \vb{\delta \mathcal{F}_{\rm mag}} \bm{\xi},
\end{equation}
where $\delta \mathcal{F}_{\rm rot}$ and $\delta \mathcal{F}_{\rm mag}$ describe the effects of rotation and an internal magnetic field, respectively. 
These effects lead to a change $\delta \omega$ in the eigenfrequency.
By describing these additional effects perturbatively, we assumed their influence to be small compared to $\vb{\mathcal{F}_0}$. 
The description and formulation of the first-order perturbative approach is given in Appendix\,\ref{App:pertub} and follows the work of \citet{loiTopologyObliquityCore2021a}.
This approach leads to the equation describing the effect of rotation and an internal magnetic field as
\begin{equation}\label{ChEq-Eq:perturbed-eigenfrequency-matrix-rot}
    \var{\omega^2} \vb{r} = \vb{M}_{\rm rot} \vb{r} + \bm{D} \vb{M}_{\rm mag} \bm{D}^\intercal \vb{r}.
\end{equation}
When the rotation axis and magnetic field axis are aligned, the Wigner $d$-matrix $\bm{D}$ is the unit matrix and we obtain $2l+1$ frequencies. 
If the axes are misaligned, the matrix $ \vb{M}_{\rm rot} + \bm{D} \vb{M}_{\rm mag} \bm{D}^\intercal$ is non-diagonal and will generally lead to $(2l+1)^2$ frequencies of which groups of $2l+1$ frequencies form an independent solution. 
Each independent solution consisting of $2l+1$ frequencies describes a single mode of the star. 
For this solution to hold, the relative amplitudes must be fixed and can be calculated using Eq.\,\eqref{ChEq-Eq:perturbed-eigenfrequency-matrix-rot}, where they are the elements of the eigenvector $\vb{r}$.
We refer to \citet{loiTopologyObliquityCore2021a} and \citet{Vandersnickt2025} for an in-depth
description and discussion of the mathematical magneto-rotational pulsation model.

To include the amplitudes in the fitting of the modes, we considered the inclination angle $i$ between the rotational axis and the line of sight and only considered 
the geometric effects \citep{gizonDeterminingInclinationRotation2003}.
For a brightness variations $I$ on the stellar surface following \vspace{-2pt}
\begin{equation}
    I (t, \theta, \phi) = \sum_{n, l, m} A_{nlm} Y^{m}_{l}(\theta, \phi) e^{i \omega_{nlm} t}, \vspace{-5pt}
\end{equation}
the amplitude correction factor is $\sqrt{\mathcal{E}(i)}$ given by \citep{dziembowskiLightRadialVelocity1977,toutainVisibilitySolarPmodes1993}
\begin{equation}
    \mathcal{E}(i) = \frac{(l - |m|)!}{(l + |m|)!} \left[ P^{|m|}_l ( \sin i)\right]^2. \vspace{-2pt}
\label{hoek}
\end{equation} 
This framework allows for the simultaneous fitting of both frequency and amplitude.
Throughout this work, we use a reference frame with polar axis equal to the rotation axis of the star.

We considered a fossil magnetic field topology as described in \citet{duezRelaxedEquilibriumConfigurations2010}, where we parametrized the field by its maximum magnetic field strength $B_{\rm max}$  and used the solutions with the largest radial scales.
By assuming a fixed topology, we reduce the number of free parameters considerably. 
We minimize the impact of the chosen topology on the results by modelling the modes independently as different modes have different sensitivity kernels in the stellar interior. 
The magnetic field was constrained to the radiative envelope of the star in a dipole-like configuration having a poloidal and a toroidal component. 
As the magnetic field topology is 
assumed to be axisymmetric along the magnetic axis,
we extended the calculation of the elements of $\vb{M}_{\rm mag}$ and corresponding eigenfunctions
to 2 dimensions. 
Appendix\,\ref{App:sensitivity} discusses the sensitivity of the modes on the selected magnetic field geometry.

\vspace{-9pt}
\section{New modelling results}\label{sec:results}
\vspace{-3pt}
\citet[][their Table\,2]{Vanlaer2025} revisited the asteroseismic modelling of HD\,192575 and determined six sets of acceptable models with different identifications for four multiplets.
We relied on these models in all the sets but for the sake of brevity we only report results from Set\,5.
The full results are available in \citet{Vandersnickt2025} and summarised in Table\,\ref{ChFit-tab:best_model_gmode}.
Our novel discovery and modelling results do not depend on the chosen set.

For each equilibrium model in the sets, we calculated the eigenfrequencies\,$(\omega_0)$ and eigenfunctions $(\bm{\xi})$ in absence of rotation and internal magnetic fields using \texttt{GYRE} \citep{goldsteinContourMethodNew2020}. 
For each mode, we subsequently calculated the change in frequency and the relative amplitudes using Eq.\,\eqref{ChEq-Eq:perturbed-eigenfrequency-matrix-rot}.
This was done using an iterative process, starting from a wide range of magnetic field strengths, rotational frequencies, and obliquity angles
and refining as needed.

Extensive testing showed proper convergence, where a modified $\chi^2$-type metric, $d^2$, defined as \vspace{-2pt}
\begin{equation}
    d^2 = \sum_{j = 1}^n \left( \frac{\Delta f_{j, {\rm theo}} - \Delta f_{j, {\rm obs}}}{\sigma_{\Delta f_j, {\rm obs}}}\right)^2 + c \cdot \sum_{j = 0}^N \left( \frac{A_{j, {\rm theo}} - A_{j, {\rm obs}}}{\sigma_{A_j, {\rm obs}}} \right)^2,
\end{equation} \vspace{-2pt}
was used, where the sum included the $n$ observed frequencies and the $N \geq n$ predicted frequencies among all independent predictions in the multiplet to explain the 9 observed frequencies and 
$\Delta\,f_i\equiv\,f_i-f_{\rm cen}$ for a chosen central frequency $f_{\rm cen}$ with selected azimuthal order $m$.
We selected $f^5_g$ as the central frequency and assume an azimuthal order $m=0$. This choice is justified as there are frequencies present for each azimuthal order in the rotational splitting, which suggests $f^5_g$ and $f^6_g$ to have azimuthal order $m=0$.
Correspondingly, we consider only the theoretical frequency differences relative to a predicted frequency with $m=0$, drastically decreasing computation time by limiting the azimuthal identification of the different frequencies.
The weight $c$ is a free parameter to fit the observed amplitudes.
The difference in summation between the frequencies and amplitudes is to penalize the fit for predicted signals that are not observed. 
For the predicted amplitudes with no observed counterpart, we set $A_{i, {\rm obs}} = 0$\,$\mu$mag and $\sigma_{A_i, {\rm obs}} = 1.6$\,$\mu$mag, which is the uncertainty on the extracted amplitudes according to the red noise corrected formulae of \citet{Montgomery1999}.
Using this metric, we calculated the corresponding confidence intervals using the F-statistic. 
We expand upon this method in Appendix\,\ref{app:likelihood}.

\begin{figure}
    \centering
         \includegraphics[width=0.85\linewidth]{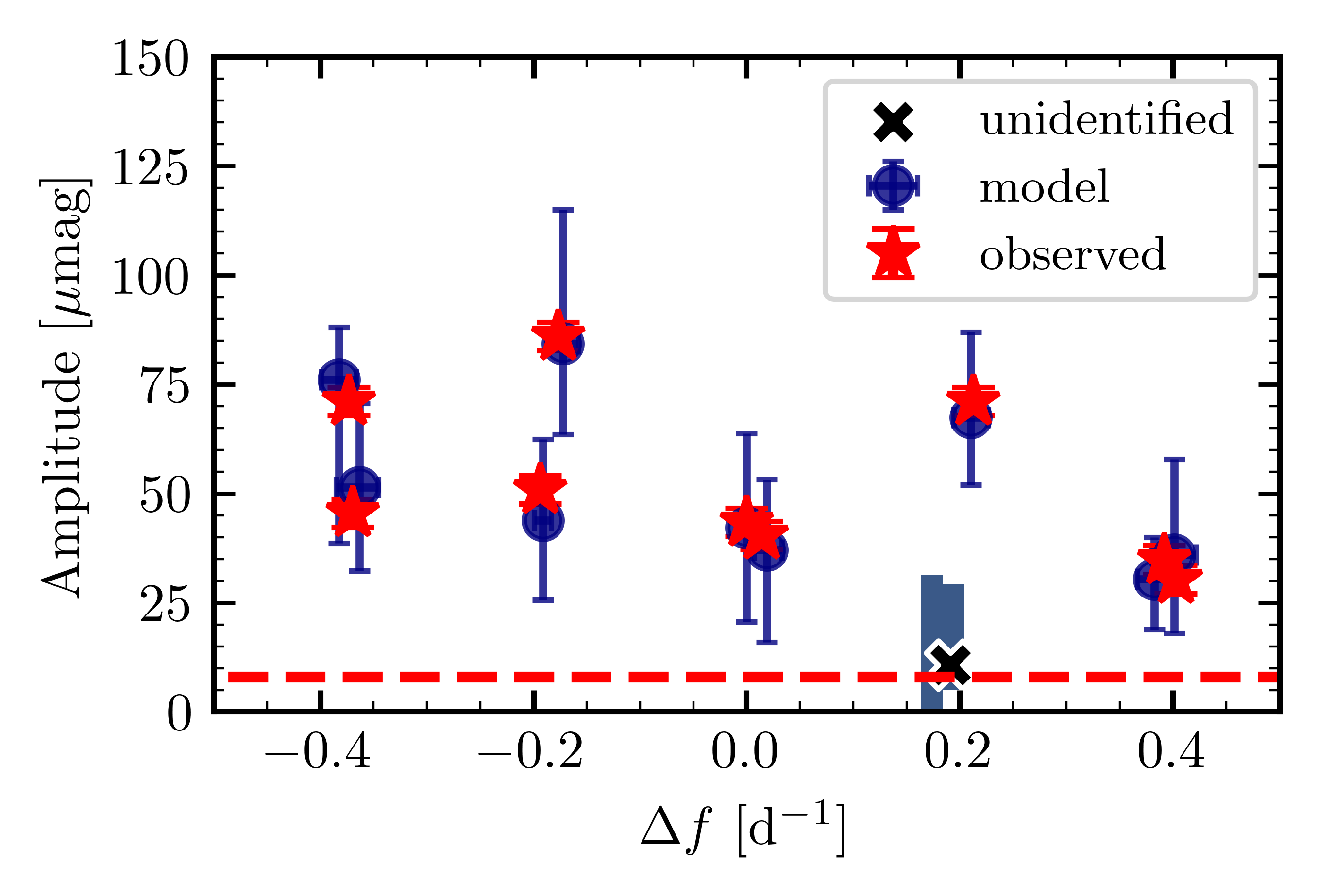}\vspace{-0.5cm}
         \label{fig:y equals x}
     \caption{Predictions of the best fitting magneto-gravito pulsation model from Set\,5 and its uncertainties (dark blue). The black cross designates a predicted frequency that is not detected in the data. The blue shaded area shows all possible locations of unidentified frequencies. The horizontal red dashed line shows the local average noise level in the data. 
     The 95\% confidence intervals for the predicted frequency differences are smaller than the size of the markers. \vspace{-10pt}
     }
     \label{fig-results:best_fit}
\end{figure}

We applied our fitting process to the split g-mode in Fig.\,\ref{ChIn:fig-HD192575-lightcurve}, which was identified to be a g$_1$ (Sets\,1 and 2) or g$_2$ mode (Sets 3 to 6) by \citet{Vanlaer2025}. In doing so we
set the fitting weight parameter to $c = 10^4$. 
This value is necessary to avoid overfitting in frequencies as the different relative uncertainties between the frequencies and amplitudes would otherwise lead to solely frequency optimization in the $(2l+1)^2$ available frequencies.
Our new magneto-gravito-asteroseismic modelling can explain all the observed frequencies in the multiplet, as illustrated for a model from Set\,5 in Fig.\,\ref{fig-results:best_fit}.
The predicted frequencies show a good agreement with the observed frequencies and amplitudes. 
As there were only 9 observed frequencies and our best pulsation model has a minimum of 10 predicted frequencies when two independent $(2l+1)$ solutions are considered,
an explanation of at least one missing frequency was required.
The black cross in Fig.\,\ref{fig-results:best_fit} shows this unidentified frequency for the best fitting model.
This frequency and its uncertainty is mostly low in amplitude and close to the mean noise level of the periodogram. 
The best fitting solution for Set 5 has a maximum magnetic field strength of $24_{-15}^{+651}$\,kG.
While the uncertainty on this value is large, the detection of the internal magnetic field follows from the detection of 9 frequencies for the $l=2$ multiplet in the updated frequency list.
We calculated the sensitivity kernels of the best fitting mode following \citet{dasSensitivityKernelsInferring2020} and found the sensitivity to peak at $R_{\rm peak} = 0.994$\,$R_{\rm star}$, indicating that the mode mostly probes the magnetic field in the near-surface region. The mean magnetic field strength at that radius is $B_{\rm peak} = 6.85$\,kG. 
The large uncertainty on $B_{\rm max}$ is connected to the large diversity of magnetic kernel sensitivities for the models in Set\,5, with a significant fraction of models probing the near-core magnetic field.
Following \citet{fullerAsteroseismologyCanReveal2015} and \citet{2016Natur.529..364S}, the quadrupole mode can only propagate in the magnetic interior if the magnetic field strength is below the critical value $B_{\rm crit}$, which ranges from $\sim\! 10^7$\,kG near the core to $\sim\! 10^2$\,kG near the surface, well above the magentic field strengths of all accepted models in the 95\% confidence interval.
This is an additional validation of the perturbative approach. 
It reinforces the importance of the Coriolis force and the value of including the centrifugal force in furture modelling of the modes.

The obliquity angle $\beta$ and the inclination angle $i$ are consistent between the different model sets, with Set 5 giving $\beta = {65^\circ} ^{+24}_{-25}$ and $i = {24^\circ}_{-17}^{+26}$.
The inclination angle has no dependence on the radial order of the mode, and only depends on the azimuthal order $m$ of the frequencies and the angular degree $l$ of the mode. 
The value of $\beta$ is determined by the small splitting and relative amplitudes, while the inclination angle is sensitive to the relative amplitudes between the frequencies.

The rotational frequency estimation of $f_{\rm rot} = 0.19_{-0.01}^{+0.03}$\,d$^{-1}$ corresponds closely to the results of \citet{Vanlaer2025}. 
We additionally computed the projected rotational velocity 
$v\sin i$ of the star. This includes the uncertainties on the inclination angle $i$, the rotational frequency $f_{\rm rot}$, and the stellar radius. The outcome is consistent with the spectroscopic result of  $27_{-8}^{+6}$\,km\,s$^{-1}$.
We find the near-core rotational velocity of the star to be around $\sim\!20$\% of the Kepler critical velocity, in line with the results of \citet{AertsTkachenko2024}.
All the best fitting model parameters are listed in Table~\ref{ChFit-tab:best_model_gmode} with the 95\% confidence intervals per model set. We find that the stellar parameter estimates based on our magnetic description are in agreement with the earlier modelling results by \citet{burssensCalibrationPointStellar2023b} and \citet{Vanlaer2025} based on a rotating non-magnetic star.

\vspace{-9pt}
\section{Conclusions}
\vspace{-3pt}
In this Letter, we presented the asteroseismic detection of an internal magnetic field inside a massive pulsator. We explained the 9 observed frequency components of the quadrupole g-mode of the $\beta\,$Cep star HD\,192575 as the combined effects of an internal magnetic field and rotation, where the magnetic and rotation axes are inclined with each other. The detected 9-frequency signature of this l=2 mode cannot be explained by 
any currently available
pulsation theory for 
low-order low-degree ($l<3$) modes in a single star without invoking an internal magnetic field. 

We focused our novel modelling work on the only multiplet detected in the TESS light curve of the star having more than the usual $2l+1$ components induced by internal rotation, after excluding the possibility of an avoided crossing between low-order modes. The star also reveals several p-mode multiplets which are well explained by the forward models used here when ignoring the magnetic field. 
We did not focus on these multiplets to assess the internal magnetic field because the theory we developed is not optimal to model p-modes with rotation kernels peaking in a stellar envelope deformed by the rotation of the star. Indeed, as shown by \citet{Vanlaer2025}, the centrifugal force needs to be included for a proper description of the low-order p-modes of HD\,192575. This requires a more complex theoretical description of the rotation based on coupling between modes of different degrees, which is currently under construction. Nevertheless, we already applied our perturbative 
magnetic 
pulsation model to the quadrupole p-mode centered at 6.5\,d$^{-1}$ as a sanity check. The results presented in Appendix\,\ref{app:results_sets} show that such simplified magneto-acoustic modelling is able to explain the four frequencies in this multiplet, with stellar parameters and rotation in agreement with those found from the quadrupole g-mode multiplet.
This good result for the p-mode encourages us to extend the current theoretical pulsation model by including the centrifugal deformation. Such novel theoretical development is underway and will be used to improve the modelling presented here.

\begin{acknowledgements} 
We thank the referee for valuable comments which helped us to improved our paper. 
The research leading to these results has received funding from the European Research Council (ERC) under the Horizon Europe programme (Synergy Grant agreement N$^\circ$101071505: 4D-STAR).  While funded by the European Union, views and opinions expressed are however those of the author(s) only and do not necessarily reflect those of the European Union or the European Research Council. Neither the European Union nor the granting authority can be held responsible for them.
VV and MV additionally acknowledge support from the Research Foundation Flanders (FWO) under grant agreement N°1156923N (PhD Fellowship) and from the KU\,Leuven Research Council (doctoral mandate grant DB/24/008), respectively.
  The TESS data presented in this paper were obtained from the Mikulski Archive for Space Telescopes (MAST) at the Space Telescope Science Institute (STScI), which is operated by the Association of Universities for Research in Astronomy, Inc., under NASA contract NAS5-26555. Support to MAST for these data is provided by the NASA Office of Space Science via grant NAG5-7584 and by other grants and contracts. Funding for the TESS mission was provided by the NASA Explorer Program.
\end{acknowledgements}


\vspace{-20pt}

%
\bibliographystyle{aa} 
\bibliography{refs} 
%

\begin{appendix}
\onecolumn
\section{Relevant frequencies of HD\,192575}
\begin{figure*}[h!]
    \centering
   \includegraphics[width=0.75\linewidth]{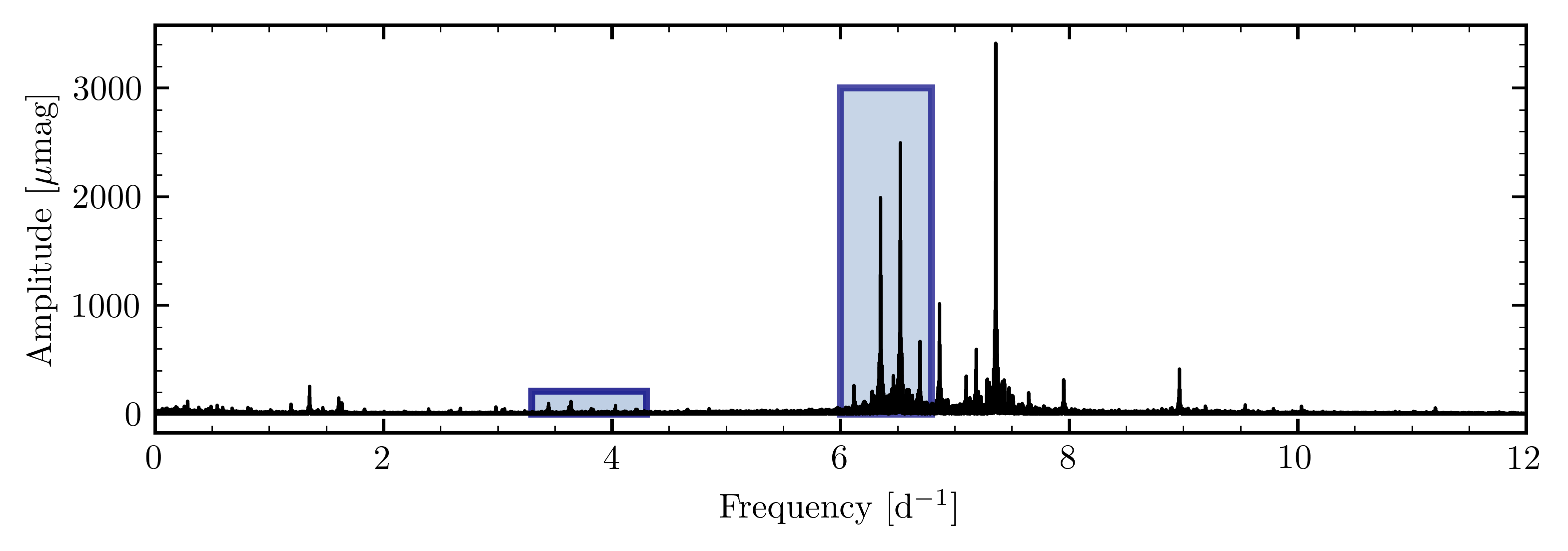}\vspace{-0.5cm}
    \caption{Lomb-Scargle periodogram of the TESS Cycle\,2,4,5, and 6 light curve of HD\,192575.
              The indicated areas are discussed in this Letter.}
    \label{ChIn:fig-HD192575-lightcurveApp}
\end{figure*}
\vspace{-10pt}
\begin{figure}[h!]
    \centering
    \begin{subfigure}[b]{0.5\textwidth}
        \includegraphics[width=0.85\textwidth]{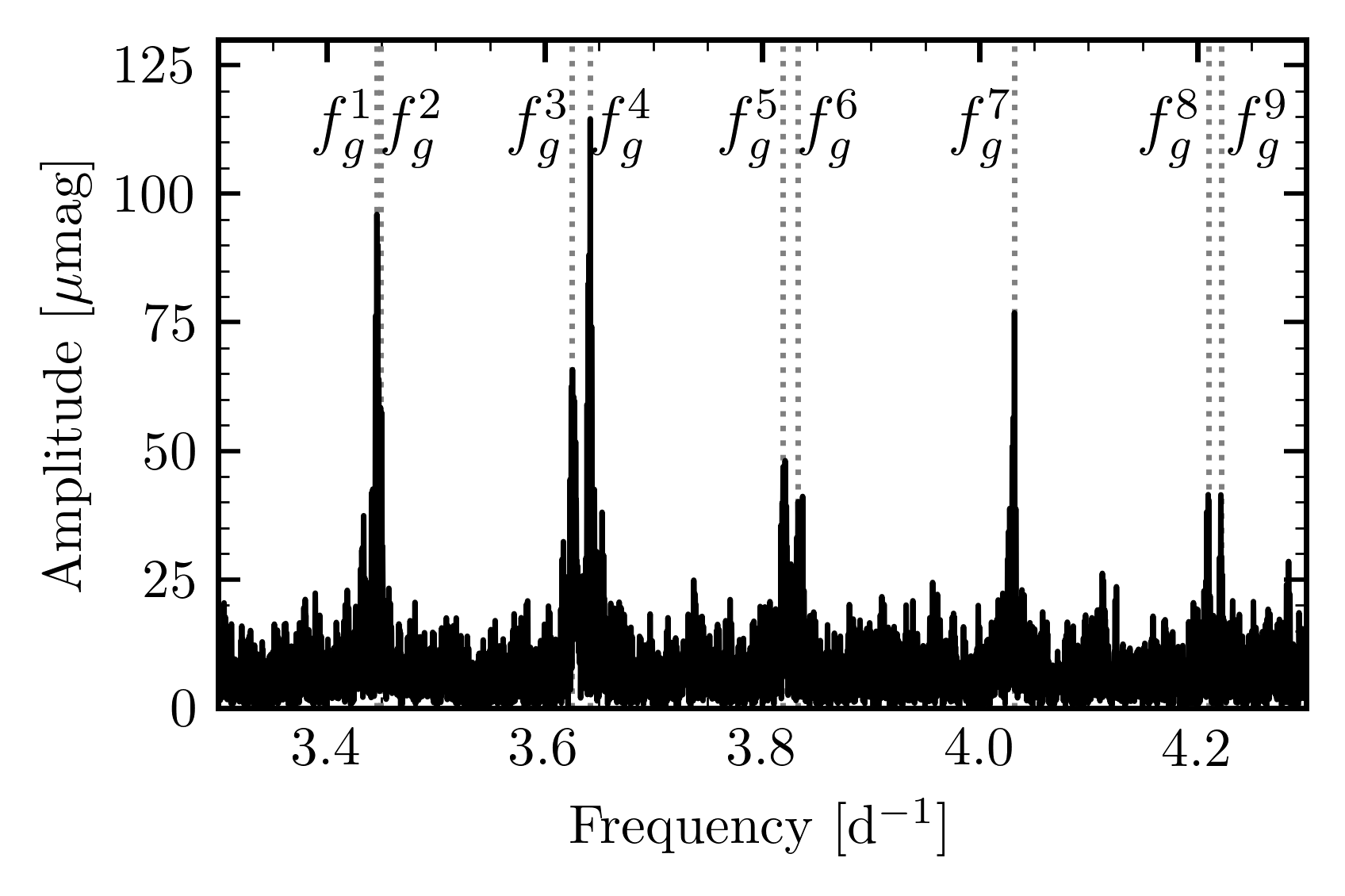} 
    \end{subfigure}%
    \begin{subfigure}[b]{0.5\textwidth}
        \includegraphics[width=0.85\textwidth]{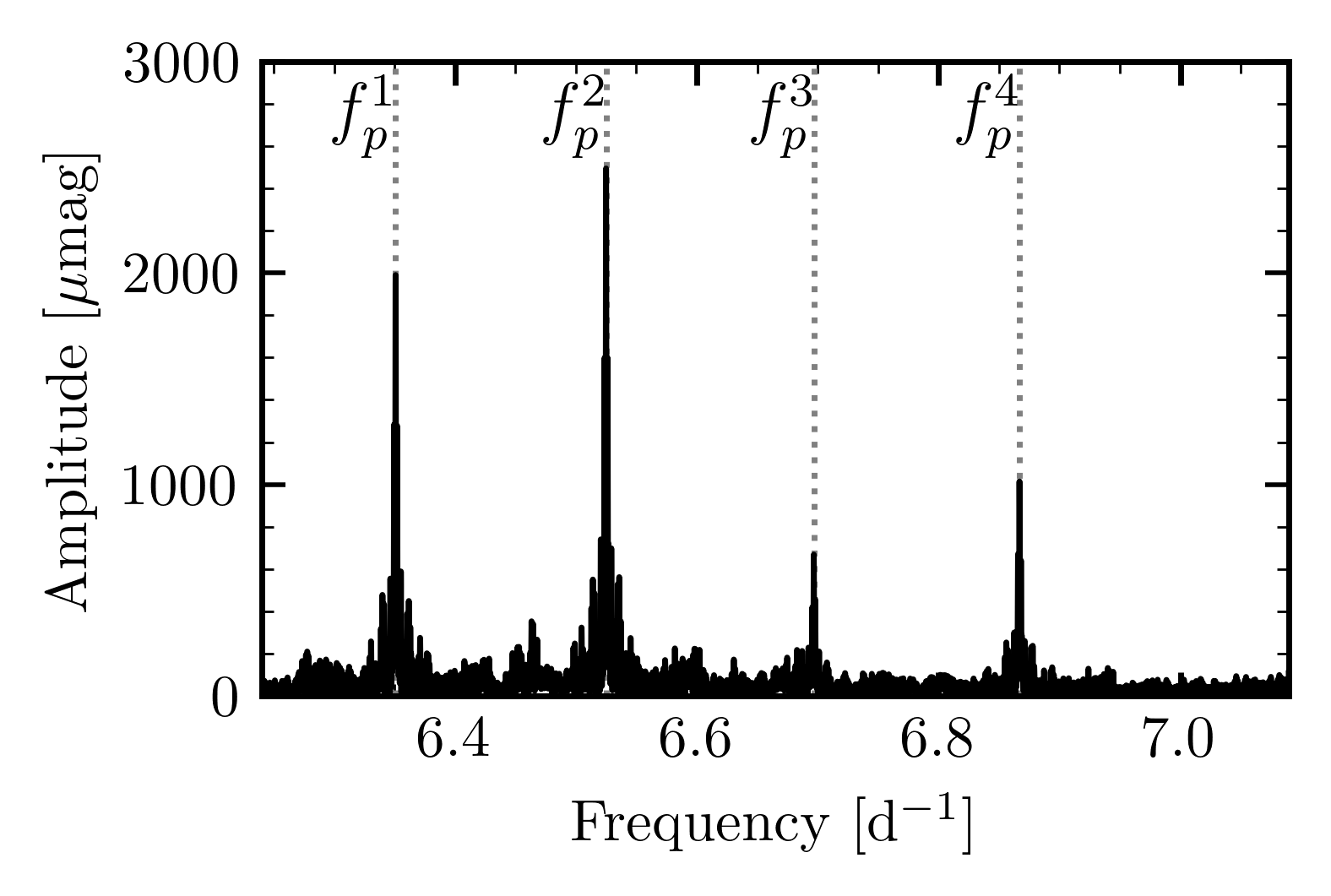} 
    \end{subfigure}\vspace{-0.5cm}
    \caption{Excerpt of the Lomb-Scargle periodogram of HD\,192575 zoomed in on the g-mode (left) and p-mode (right) multiplet under study.
    The dotted lines indicate the frequencies extracted from the data and the mode labels refer to Table\,\ref{ChFit-tab:fitting_results}.}
    \label{ChIn:fig-HD192575-lightcurve}
\end{figure}
\vspace{-10pt}
\begin{table}[h!]
    \centering
    \caption{Best stellar parameters for the different sets of stellar models used in this work, as determined by \citet{Vanlaer2025}. }
    
    \begin{tabular}{l|cccccc}
        \hline
        \hline
        Parameter & Set 1 & Set 2 & Set 3 & Set 4 & Set 5 & Set 6 \\
        \hline
        $M_{\rm ini}$ [M$_\odot$]   & $13.0_{-2.5}^{+0.5}$ & $11.5_{-1.0}^{+2.0}$ & $11.0_{-2.0}^{+1.0}$  & $12.5_{-2.0}^{+1.0}$  & $12.0_{-1.5}^{+1.5}$ & $12.0_{-1.5}^{+1.5}$\\
        $Z_{\rm ini}$ [dex]         & $0.014_{-0.002}^{+0.002} $ & $0.012_{-0.000}^{+0.004}$   & $0.016_{-0.004}^{+0.000}$   & $0.014_{-0.002}^{+0.002} $   & $0.014_{-0.002}^{+0.002} $  & $ 0.012_{-0.000}^{+0.004}$  \\ 
        $f_{\rm CBM}$               & $0.035_{-0.030}^{+0.000}$ & $0.035_{-0.030}^{+0.000} $ & $0.025_{-0.020}^{+0.010} $ & $0.010_{-0.005}^{+0.025}$  & $0.10_{-0.005}^{+0.025} $ & $0.015_{-0.010}^{+0.020} $ \\
        $X_c$                       & $0.236_{-0.220}^{+0.025}$ & $0.136_{-0.120}^{+0.175}$  & $0.251_{-0.235}^{+0.160}$  & $0.016_{-0.000}^{+0.285}$  & $0.026_{-0.010}^{+0.225}$ & $0.016_{-0.000}^{+0.395}$  \\ 
        \hline
        g-mode      & $(2, -1)$ & $(2, -1)$ & $(2, -2)$ & $(2, -2)$ & $(2, -2)$ & $(2, -2)$ \\
        p-mode      & $(2, +2)$ & $(2, +3)$ & $(2, +0)$ & $(2, +1)$ & $(2, +2)$ & $(2, +3)$ \\
        \hline
    \end{tabular}
    \label{ChFit-tab:stellar_models}
\end{table}

\begin{table*}[h!]
    \centering
    \caption{Extracted frequencies, amplitudes, and SNR of the quadrupole g- and p-mode under study.}
    \begin{tabular}{l|ccc || l|ccc}
        \hline
        \hline
         & frequency [d$^{-1}$]  & amplitude [$\mu$mag] & SNR &  & frequency [d$^{-1}$]  & amplitude [$\mu$mag] & SNR \\ 
         \hline
         $f^1_g$ & 3.445953(6)  & 71.2(1.6) & 17.3 & $f^1_p$ & 6.3509313(2) & 1835.6(1.6) & 702.8 \\
         $f^2_g$ & 3.449320(10) & 45.5(1.6) & 11.0 & $f^2_p$ & 6.5250052(2) & 2331.9(1.6) & 878.1 \\
         $f^3_g$ & 3.625411(9)  & 50.9(1.6) & 12.3 & $f^3_p$ & 6.6968226(6) & 731.8(1.6)  & 268.3 \\
         $f^4_g$ & 3.642091(5)  & 86.0(1.6) & 20.9 & $f^4_p$ & 6.8670011(5) & 952.8(1.6)  & 337.4 \\
         $f^5_g$ & 3.819262(11) & 43.4(1.6) & 10.7 &  & & & \\
         $f^6_g$ & 3.833089(11) & 40.4(1.6) & 10.1 &  & & & \\
         $f^7_g$ & 4.032025(6)  & 71.1(1.6) & 19.6 &  & & & \\
         $f^8_g$ & 4.210734(13) & 34.8(1.6) & 10.4 &  & & & \\
         $f^9_g$ & 4.221790(15) & 30.3(1.6) & 9.1  &  & & & \\
         \hline
    \end{tabular}
    \label{ChFit-tab:fitting_results}
    \tablefoot{The full, electronic table including all extracted signals, including uncertainties and SNR, is available at \url{https://github.com/Mathijs-Vanrespaille/Vandersnickt_2025/blob/main/table_A1_complete.txt}.}
\end{table*}

\section{Kernel sensitivity}\label{App:sensitivity}

\begin{figure*}[h!]
    \centering
   \includegraphics[width=0.9\linewidth]{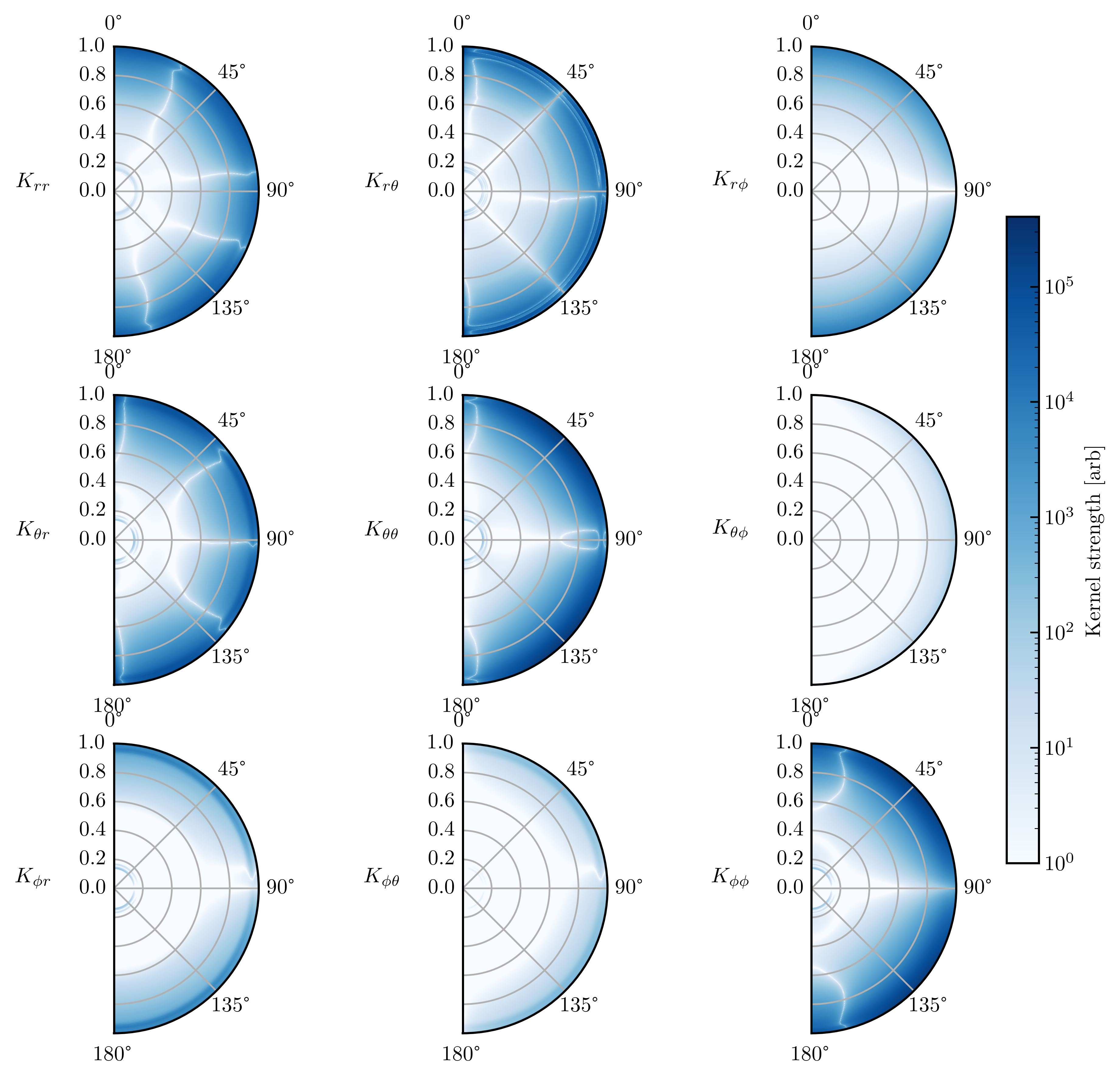}\vspace{-0.5cm}
    \caption{Magnetic sensitivity kernel of an example g mode with $n=-1$, $l=2$.}
    \label{fig:mag_sens_kernel}
\end{figure*}

Following the results of \citet{dasSensitivityKernelsInferring2020}, the sensitivity of the modes to the internal magnetic field can be decomposed as a product of the magnetic stress tensor $\bm{BB}$ and the sensitivity kernel $\bm{K}$ as 
\begin{equation}
    \langle \delta \mathcal{F}(\bm{\xi}), \bm{\xi}\rangle = \int_V \bm{BB}:\bm{K} \text{d} V,
\end{equation}
where ":" denotes the tensor product. 
Figure\,\ref{fig:mag_sens_kernel} shows an example of the magnetic sensitivity kernel. The sensitivity peak is near the surface for this example and varies multiple order of magnitudes throughout the interior. Other cases are possible where the sensitivity is strongly peaked near the core, leading to higher magnetic field strengths required for a similar effect on the modes.

\begin{figure*}[h!]
    \centering
   \includegraphics[width=0.9\linewidth]{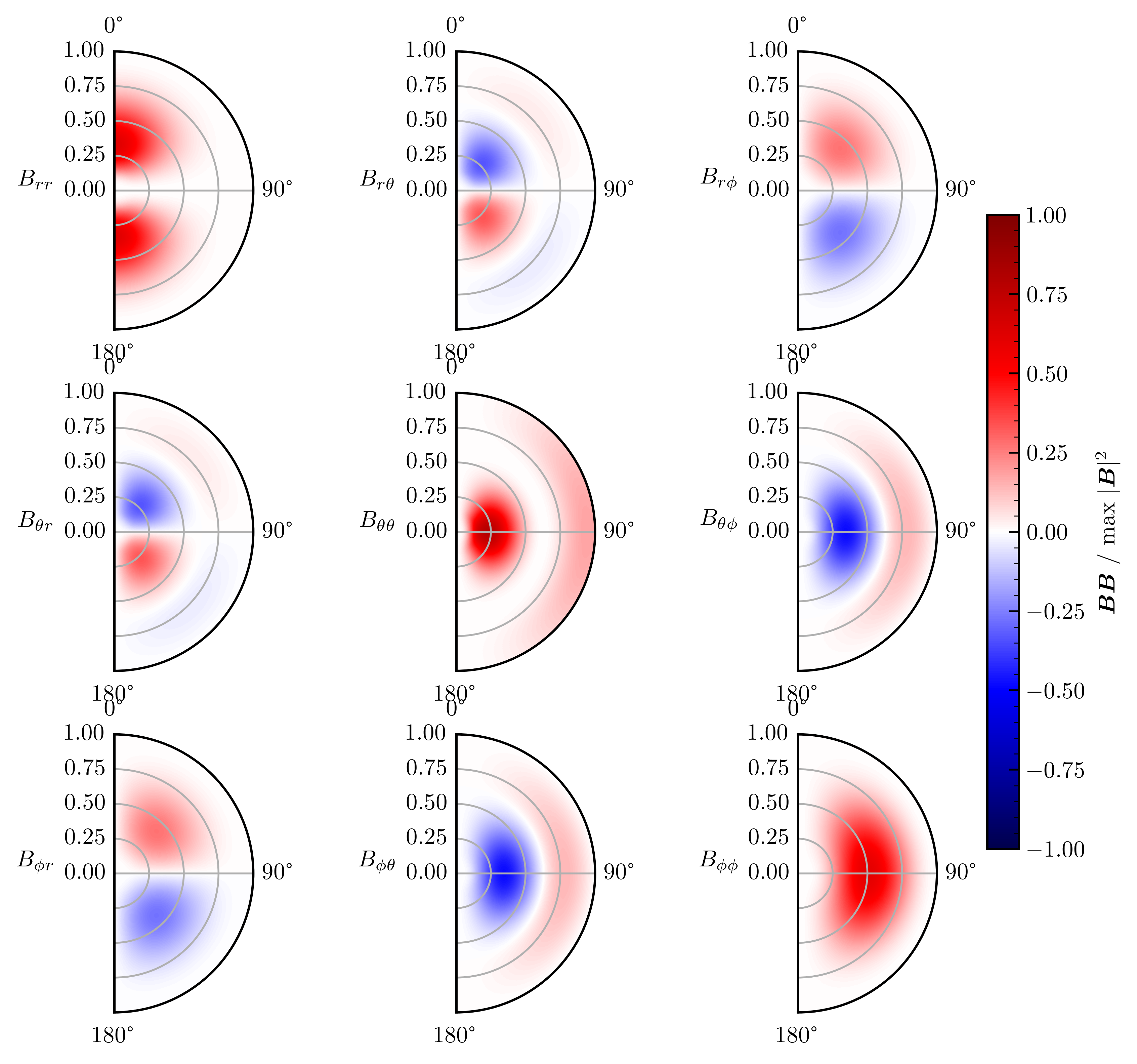}\vspace{-0.5cm}
    \caption{Magnetic stress tensor of an example magnetic field used following the prescription of \citet{duezRelaxedEquilibriumConfigurations2010}.}
    \label{fig:magnetic_stress_tensor}
\end{figure*}

Figure\,\ref{fig:magnetic_stress_tensor} shows the magnetic stress tensor for an example magnetic field topology used. These figures show that the maximum magnetic field value is not at the edges of the field. The magnetic field strength peak is, however, deep in the envelope close to the core.

\twocolumn

\section{Perturbative formulation}\label{App:pertub}
The perturbation as a consequence of rotation is given by
\begin{equation}
    \delta \vb{\mathcal{F}_{\rm rot}} = - 2 i \omega_0 \vb{\Omega} \times \bm{\xi},
\end{equation}
where $\vb{\Omega}$ is the rotation vector of the star. 

The magnetic perturbation operator for an internal magnetic field $\vb{B_0}$ is given by 
\begin{equation}
\begin{aligned}
    \delta \vb{\mathcal{F}_{\rm mag}}\ \bm{\xi} & =   \frac{1}{\rho_0}\left[\vb{B_0} \times \left[ \grad \times \left( - \vb{B_0}(\grad \cdot \bm{\xi}) + (\vb{B_0} \cdot \grad)\bm{\xi} - (\bm{\xi} \cdot \grad)\vb{B_0} \right) \right] \right] \\
    & + \frac{1}{\rho_0}\left[(\grad \times (\bm{\xi} \times \vb{B_0}))\times (\grad \times \vb{B_0})  \right]  
                                                 - \frac{\rho'}{\rho_0^2}(\grad \times \vb{B_0})\times \vb{B_0}.
\end{aligned}
\end{equation}
These first order approximations require the rotational frequency $\Omega \ll \omega_0$ and the Alvén wave frequency $\omega_{\rm A} \ll \omega_0$ and additionally ignore any deformation of the star.

We allowed for an obliquity angle $\beta$ between the rotation axis and the magnetic field axis. 
The change in the eigenfrequency is consequently given by 
\begin{equation}\label{ChEq-Eq:perturbed-eigenfrequency-matrix-rot}
    \var{\omega^2} \vb{r} = \vb{M}_{\rm rot} \vb{r} + \bm{D} \vb{M}_{\rm mag} \bm{D}^\intercal \vb{r},
\end{equation}
where we have
\begin{align}
    \vb{M}_{\rm rot} &= \frac{\text{diag}\left(  \langle \var{\mathcal{F}_{\rm rot}}(\bm{\xi_{\rm rot}^{(m)}}), \bm{\xi_{\rm rot}^{(m)}}\rangle  | \text{ for } m = -l, \cdots, l \right)}{\langle \bm{\xi_0}, \bm{\xi_0}\rangle}, \label{ChEq-Eq:rotationMatrix}\\
    \vb{M}_{\rm mag} &= \frac{\text{diag}\left(  \langle \var{\mathcal{F}_{\rm mag}}(\bm{\xi_{\rm mag}^{(m)}}), \bm{\xi_{\rm mag}^{(m)}}\rangle  | \text{ for } m = -l, \cdots, l \right)}{\langle \bm{\xi_0}, \bm{\xi_0}\rangle} \label{ChEq-Eq:magneticMatrix}
\end{align}
with $l$ the angular degree and $m$ the azimuthal order and where $\bm{D}(\beta)$ is the Wigner $d$-matrix describing the change in frame between the oblique rotation and magnetic field frames, such that each matrix $\vb{M}$ is calculated in its respective reference frame.
$\langle . , . \rangle $ denotes the inner product of the eigenvectors $\vb{\xi}$ over the star.
Equation \eqref{ChEq-Eq:perturbed-eigenfrequency-matrix-rot} is obtained by projecting the magnetic modes to the corotating frame.

\section{Calculating the likelihood function confidence regions}\label{app:likelihood}

For each combination of parameters, we obtain a distance $d^2$ between the observed and predicted frequencies and amplitudes.
The best fitting model minimizes this distance to $d^2_{\rm min}$. 
For each model, we can calculate the F-statistic to compare with the best fit. 
This statistic is a measure of the difference in the explained variance expressed by the distance and is given by
\begin{equation}
    \label{ChFit:eq:F-statistic}
    {\rm F} (d^2) = \frac{d^2 - d^2_{\rm min}}{d^2_{\rm min}} \frac{N-K}{K} \sim F_{K, N-K},
\end{equation}
for a number of observables $N$, and a number of free parameters $K$.
The F-statistic is distributed as an $F$-distribution with $K$ and $N-K$ degrees of freedom.
This leads to an asymptotic $100(1-\alpha)$\% confidence region of the parameters $\bm{\theta}$, given by
\begin{equation}
    \left\{ \bm{\theta} \bigg| d^2(\bm{\theta}) \leq d^2_{\rm min} \left( 1 + \frac{K}{N-K}F_\alpha \right)\right\}.
\end{equation}
We take $K=4$ to estimate the parameters connected to rotation and the internal magnetic field. 
The uncertainty intervals in Tables\,\ref{ChFit-tab:best_model_gmode} and  \ref{ChFit-tab:best_model_pmode} show these parameters of the magnetic field and rotation. 
The uncertainties on the other parameters can be interpreted as the interval necessary to obtain the corresponding uncertainty in the magnetic field and rotation parameters. 

Equation~\ref{ChFit:eq:F-statistic} gives the expected distribution of the F-statistic based on the metric $d^2$. 
We can use this distribution to define a likelihood function for a set of parameters $\bm{\theta}$, given by
\begin{equation}
    \mathcal{L}(\bm{\theta}) = F_{K, N-K} \left( \frac{d^2(\bm{\theta}) - d^2_{\rm min}}{d^2_{\rm min}} \frac{N-K}{K} \right).
\end{equation}

\section{Results for each model set for both quadrupole modes}\label{app:results_sets}

The fitting process of the p-mode centered on 6.5\,d$^{-1}$ shown in Figs\,\ref{ChIn:fig-HD192575-lightcurve} and \ref{ChIn:fig-HD192575-lightcurveApp} differs slightly from the one for the g-mode, as we can only consider four observed frequencies in the multiplet. We identified $f^3_p$ as the central frequency $f_{\rm cen}$ by comparing the relative amplitudes to the g-mode. We additionally required the predicted set of frequencies to consist of a single independent magnetic solution, as no clear splittings aside from those expected from rotation are available. The weight parameter was set to $c=10^3$.

\begin{figure}
    \centering
       \includegraphics[width=0.8\linewidth]{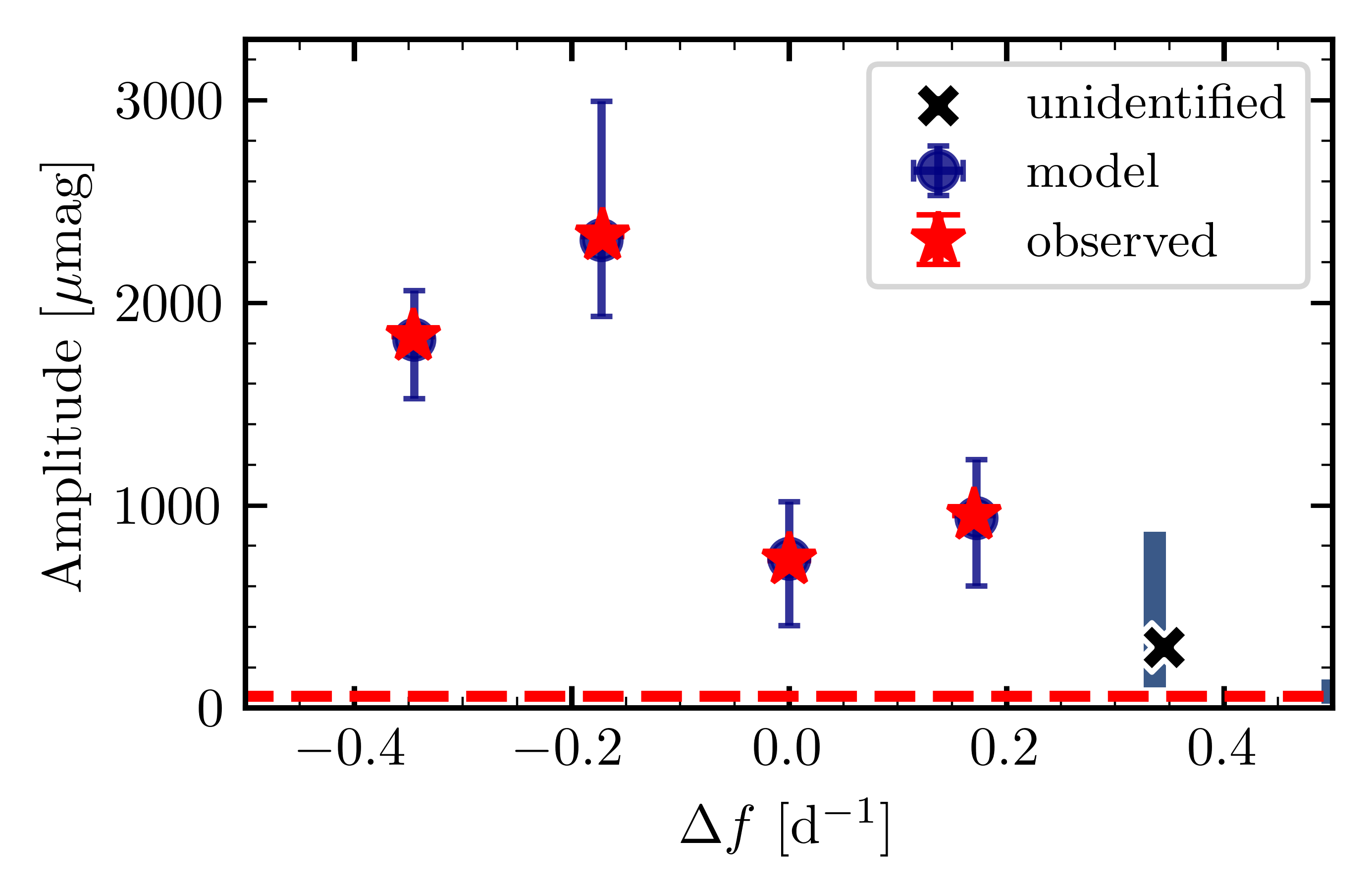}\vspace{-0.5cm}
     \caption{As Fig.\,\ref{fig-results:best_fit} but for the four observed components of the quadrupole p-mode.}
     \label{fig-results:best_fit2}
\end{figure}

Figure\,\ref{fig-results:best_fit2} shows the best fitting frequencies and amplitudes for the models in Set\,5.
There is a good agreement between the observed and predicted frequencies.  The best fitting model predicts an additional frequency that is not identified in the observed frequencies, but it has a low amplitude.
The narrow frequency confidence interval on the predicted frequencies can be immediately connected to the low uncertainty on the predicted rotational frequency reported in Table~\ref{ChFit-tab:best_model_pmode}, as this factor has the largest and most direct impact on the predicted frequencies. 
All predicted parameters are similar to the values derived from the g-mode fittings described in the main text. 

The probing depth of the kernel for the best model in Set\,5 peaks at $R_{\rm peak}=0.996$\,$R_{\rm star}$ with a magnetic field strength average of $B_{\rm peak} = 1.86$\,kG at that radius.
The maximum field strength is $B_{\rm max} = 6.5_{-5.1}^{+20.0}$\,kG. 
The current framework ignores the centrifugal force, the deformation it induces in the stellar envelope, and the accompanying couplings between spheroidal and toroidal components caused by that force.
In addition to this omission, the internal magnetic field could also be more complex than a dipole-like field.
These effects will be especially important to model p-mode splittings for HD\,192575, which has been classified as a moderate rotator by \citet{AertsTkachenko2024}.

The predicted projected rotational velocity $v\,\sin i$ is again consistent with the previous estimate by \citet{burssensCalibrationPointStellar2023b}, with a larger uncertainty due to the larger uncertainty 
on the inclination angle $i$.

\begin{table*}[h!]
    \centering
    \caption{Best model parameters and 95\% confidence intervals for the asteroseismic modelling of the g-mode.}
    \label{ChFit-tab:best_model_gmode}
    \begin{tabular}{l|cccccc}
        \hline
        \hline
        Parameter & Set 1 & Set 2 & Set 3 & Set 4 & Set 5 & Set 6 \\
        \hline
        $M$ $[{\rm M_\odot}]$   & $13.0_{-1.5}^{+0.5}$       & $12.5_{-1.0}^{+0.5}$       & $12.0_{-1.5}^{+0.5}$        & $12.5_{-1.0}^{+1.0}$      & $11.5_{-1.0}^{+2.0}$        & $12.5_{-1.0}^{+1.0}$\\
        $R$ $[{\rm R_\odot}]$   & $9.0_{-0.9}^{+1.0}$       & $10.6_{-0.7}^{+0.1}$       & $7.7_{-0.5}^{+0.3}$        & $10.8_{-2.9}^{+0.0}$      & $8.7_{-0.7}^{+2.3}$        & $10.5_{-1.5}^{+0.5}$\\
        $\log L /{\rm L_\odot}$ & $4.446_{-0.182}^{+0.002}$  & $4.435_{-0.100}^{+0.013}$ & $4.284_{-0.134}^{+0.041}$  & $4.422_{-0.197}^{+0.024}$  & $4.241_{-0.044}^{+0.207}$        & $4.410_{-0.125}^{+0.038}$\\
        $\log T_{\rm eff} / \rm{K}$ & $4.40_{-0.05}^{+0.01}$  & $4.35_{-0.01}^{+0.01}$  & $4.390_{-0.028}^{+0.005}$   & $4.351_{-0.004}^{+0.058}$  & $4.353_{-0.007}^{+0.054}$    & $4.354_{-0.008}^{+0.035}$\\
        $\log g$                    & $3.64_{-0.12}^{+0.07}$   & $3.48_{-0.01}^{+0.04}$  & $3.74_{-0.04}^{+0.03}$     & $3.47_{-0.00}^{+0.28}$      & $3.62_{-0.17}^{+0.12}$        & $3.49_{-0.04}^{+0.14}$\\
        Age [Myr]                   & $14.1_{-2.4}^{+3.2}$            & $16.0_{-1.6}^{+3.4}$          & $14.6_{-2.0}^{+5.7}$      & $16.1_{-4.7}^{+0.5}$    & $16.6_{-5.2}^{+5.2}$      & $15.9_{-3.5}^{+1.9}$ \\
        $X_c$                       & $0.231_{-0.210}^{+0.010}$  & $0.136_{-0.120}^{+0.030}$  & $0.241_{-0.020}^{+0.040}$   & $0.026_{-0.010}^{+0.150}$ & $0.056_{-0.040}^{+0.140}$   & $0.051_{-0.040}^{+0.115}$ \\
        $M_{\rm{cc}}$ $[{\rm M_\odot}]$  & $3.55_{-1.51}^{+0.03}$ & $2.96_{-0.97}^{+0.21}$   & $2.85_{-0.51}^{+0.09}$     & $2.30_{-0.31}^{+0.90}$      & $1.86_{-0.12}^{+1.20}$        & $2.37_{-0.52}^{+0.66}$\\
        $R_{\rm{cc}}$ $[{\rm R_\odot}]$ & $0.117_{-0.046}^{+0.004}$ & $0.089_{-0.022}^{+0.007}$ & $0.120_{-0.007}^{+0.004}$ & $0.0699_{-0.0007}^{+0.0449}$  & $0.080_{-0.010}^{+0.034}$     & $0.076_{-0.009}^{+0.027}$\\ 
        
        \hline
        $B_{\rm max}$ $[{\rm kG}]$    & $24_{-21}^{+48}$            & $3.2_{-0.5}^{+17.2}$       & $698_{-398}^{+865}$       & $5.1_{-0.7}^{+776.5}$       & $24_{-15}^{+651}$         & $31_{-28}^{+65}$\\
        $f_{\rm rot}$ $[$d$^{-1}$$]$  & $0.19_{-0.01}^{+0.02}$     & $0.19_{-0.01}^{+0.02}$     & $0.19_{-0.01}^{+0.02}$    & $0.19_{-0.01}^{+0.02}$    & $0.19_{-0.01}^{+0.03}$      & $0.19_{-0.01}^{+0.02}$\\
        $\beta$ $[{^\circ}]$            & $61_{-17}^{+25}$           & $63_{-22}^{+25}$           & $60_{-20}^{+28}$          & $60_{-21}^{+28}$           & $65_{-25}^{+24}  $         & $61_{-13}^{+30}$\\
        $i$ $[{^\circ}]$                & $24^{+24}_{-12} $          & $24^{+25}_{-12}$            & $24^{+24}_{-10}$            & $24^{+24}_{-10}$            & $24^{+26}_{-17}$             & $24^{+15}_{-10}$\\
        \hline
        $R_{\rm peak}$ $[R_{\rm star}]$          & $0.989$                   & $0.994$                   & $0.144$                   & $0.994$                   & $0.994$                     & $0.994$\\   
        $B_{\rm peak}$ $[{\rm kG}]$             & $6.76$                    & $0.88$                    & $248$                     & $1.44$                      & $6.85$                        & $8.75$\\
        $s = 2 f_{\rm rot} / f_{\rm cen}$              & $0.100_{-0.007}^{+0.010}$  & $0.100_{-0.007}^{+0.013}$  & $0.100_{-0.007}^{+0.009}$ & $0.100_{-0.007}^{+0.009}$& $0.100_{-0.007}^{+0.013}$ & $0.100_{-0.007}^{+0.012}$ \\
        $\Omega / \Omega_{\rm crit}^{\rm Kepler}$  & $0.16_{-0.01}^{+0.06}$     & $0.21_{-0.02}^{+0.02}$     & $0.14_{-0.01}^{+0.01}$    & $0.22_{-0.09}^{+0.01}$    & $0.17_{-0.03}^{+0.07}$      & $0.21_{-0.05}^{+0.03}$\\
        $v \sin i$   $[{\rm km/s}] $            & $35_{-32}^{+37}$          & $42_{-19}^{+35}$          & $30_{-13}^{+27}$          & $42_{-24}^{+35}$          & $34_{-21}^{+44}$            & $41_{-20}^{+38}$ \\
        \hline
    \end{tabular}
\end{table*}

\begin{table*}[h!]
    \centering
    \caption{Best model parameters and 95\% confidence intervals for the asteroseismic modelling of the p-mode.}
    \label{ChFit-tab:best_model_pmode}
    \begin{tabular}{l|cccccc}
        \hline
        \hline
        Parameter & Set 1 & Set 2 & Set 3 & Set 4 & Set 5 & Set 6 \\

        \hline
        $M$ $[{\rm M_\odot}]$   & $12.5_{-1.0}^{+1.0}$       & $12.0_{-0.5}^{+1.0}$       & $11.0_{-0.5}^{+1.5}$        & $13.5_{-2.0}^{+0.0}$      & $12.5_{-2.0}^{1.0}$        & $13.5_{-2.0}^{+0.0}$\\
        $R$ $[{\rm R_\odot}]$   & $8.4_{-0.4}^{+1.6}$        & $10.4_{-0.5}^{+0.3}$       & $7.7_{-1.1}^{+0.3}$        & $8.6_{-0.7}^{+0.2.2}$      & $8.2_{-0.2}^{+2.8}$        & $9.3_{-0.3}^{+1.7}$\\
        $\log L /{\rm L_\odot}$ & $4.362_{-0.099}^{+0.086}$  & $4.390_{-0.056}^{+0.058}$ & $4.205_{-0.056}^{+0.119}$  & $4.447_{-0.221}^{+0.000}$  & $4.321_{-0.124}^{+0.127}$        & $4.445_{-0.159}^{+0.003}$\\
        $\log T_{\rm eff} / \rm{K}$ & $4.39_{-0.04}^{+0.02}$ & $4.340_{-0.004}^{+0.018}$  & $4.370_{-0.009}^{+0.030}$   & $4.406_{-0.059}^{+0.003}$  & $4.39_{-0.04}^{+0.02}$    & $4.39_{-0.04}^{+0.00}$\\
        $\log g$                    & $3.68_{-0.16}^{+0.03}$ & $3.478_{-0.008}^{+0.041}$  & $3.708_{-0.005}^{+0.161}$     & $3.70_{-0.22}^{+0.05}$      & $3.71_{-0.25}^{+0.03}$        & $3.63_{-0.17}^{+0.00}$\\
        Age [Myr]                   & $14.2_{-2.5}^{+3.1}$     & $17.5_{-3.1}^{+1.9}$          & $18.1_{-5.4}^{+2.3}$      & $12.6_{-1.2}^{+4.0}$    & $13.2_{-11.8}^{+8.6}$      & $12.4_{-0.0}^{+5.39}$ \\
        $X_c$                       & $0.226_{-0.205}^{+0.015}$  & $0.126_{-0.110}^{+0.040}$  & $0.261_{-0.040}^{+0.080}$   & $0.191_{-0.175}^{+0.010}$ & $0.191_{-0.175}^{+0.000}$   & $0.141_{-0.125}^{+0.025}$ \\
        $M_{\rm{cc}}$ $[{\rm M_\odot}]$  & $3.12_{-1.10}^{+1.02}$ & $2.74_{-0.75}^{+0.43}$   & $2.75_{-0.41}^{+0.33}$     & $3.18_{-1.19}^{+0.01}$      & $2.69_{-0.95}^{+0.37}$        & $2.90_{-1.05}^{+0.13}$\\
        $R_{\rm{cc}}$ $[{\rm R_\odot}]$ & $0.116_{-0.045}^{+0.005}$ & $0.086_{-0.019}^{+0.428}$ & $0.118_{-0.005}^{+0.026}$ & $0.114_{-0.044}^{+0.001}$ & $0.111_{-0.041}^{+0.003}$     & $0.101_{-0.034}^{+0.001}$\\ 
        
        \hline
        $B_{\rm max}$ $[{\rm kG}]$    & $6.4_{-4.9}^{+17.1}$      & $0.86_{-0.11}^{+6.68}$       & $13_{-7}^{+66}$          & $10_{-8}^{+15}$       & $6.5_{-5.1}^{+20.0}$         & $2.3_{-1.5}^{+5.3}$\\
        $f_{\rm rot}$ $[\rm{d}^{-1}]$  & $0.172_{-0.003}^{+0.002}$ & $0.172_{-0.003}^{+0.002}$ & $0.172_{-0.003}^{+0.002}$ & $0.172_{-0.003}^{+0.002}$  & $0.172_{-0.003}^{+0.002}$    & $0.173_{-0.003}^{+0.002}$\\
        $\beta$ $[{^\circ}]$            & $67_{-35}^{+22}$          & $72_{-40}^{+17}$          & $68_{-47}^{+21}$          & $67_{-36}^{+22}$           & $67_{-36}^{+22}$             & $74_{-41}^{+15}$\\
        $i$ $[{^\circ}]$                & $27^{+30}_{-24}$          & $29^{+38}_{-17}$          & $27^{+32}_{-25}$          & $26^{+30}_{-24}$           & $27^{+28}_{-24}$             & $27^{+29}_{-15}$\\
        \hline
       $R_{\rm peak}$ $[R_{\rm star}]$          & $0.995$                   & $0.995$                   & $0.995$                   & $0.995$                   & $0.996$                     & $0.996$\\   
        $B_{\rm peak}$ $[{\rm kG}]$             & $1.81$                    & $0.23$                    & $3.67$                    & $2.97$                    & $1.87$                      & $0.65$\\
        $s = 2 f_{\rm rot} / f_{\rm cen}$              & $0.0515_{-0.0009}^{+0.0007}$  & $0.0515_{-0.0009}^{+0.0007}$  & $0.0515_{-0.0008}^{+0.0007}$ & $0.0515_{-0.0009}^{+0.0007}$& $0.0515_{-0.0009}^{+0.0007}$ & $0.0515_{-0.0009}^{+0.007}$ \\
        $\Omega/\Omega_{\rm crit}^{\rm Kepler}$  & $0.139_{-0.007}^{+0.043}$ & $0.195_{-0.016}^{+0.010}$ & $0.128_{-0.033}^{+0.003}$ & $0.139_{-0.017}^{+0.067}$  & $0.139_{-0.014}^{+0.070}$    & $0.155_{-0.003}^{+0.054}$\\
        $v \sin i$ $[{\rm km/s}] $              & $33_{-30}^{+38}$      & $29_{-26}^{+48}$          & $30_{-30}^{+25}$          & $33_{-30}^{+42}$          & $31_{-27}^{+44}$            & $37_{-34}^{+41}$ \\
        \hline
    \end{tabular}
\end{table*}

\end{appendix}

\end{document}